\documentclass[12pt]{article}
\usepackage[pctex32]{graphics}
\usepackage[pctex32]{graphics}
\begin{document}
~
\vspace{1cm}
\begin{center} {\Large \bf  Entropy and Quantum States of Tachyon Supertube}
                                                  
\vspace{1cm}

                      Wung-Hong Huang\\
                       Department of Physics\\
                       National Cheng Kung University\\
                       Tainan, Taiwan\\

\end{center}
\vspace{2cm}
\begin{center} {\large \bf  Abstract} \end{center}
We quantize the deformed modes of a single supertube solution with regular profile of circular cross section on the unstable non-BPS D3-branes by using the Minahan-Zwiebach tachyon action.  The result is used to count the microstates in an ensemble of supertube with fixed macroscopic quantities of charges $Q_{D0}$, $Q_{F1}$ and angular momentum $J$.   We show that the entropy of the system is proportional to $\sqrt{Q_{D0}Q_{F1} - J}$, which is consistent with that calculated by the DBI action.   Therefore, besides the well known properties that the kink solution (and its fluctuation) of tachyon DBI action corresponds with the brane solution (and its fluctuation) of DBI action, our result establishes a property that the entropy of the tachyon supertube in tachyonic DBI action corresponds with that of the supertube in  DBI action.

\vspace{2cm}
\begin{flushleft}
E-mail:  whhwung@mail.ncku.edu.tw\\
\end{flushleft}
\newpage
\section {Introduction}

Supertubes found by  Mateos and Townsend [1] are tubular bound states of D0-branes, fundamental strings (F1) and D2-branes, which are supported against collapse by the angular momentum generated by the Born-Infeld (BI) electric and magnetic fields.  Following the initial supertube paper, a matrix model version of it was introduced by Bak and Lee [2]. A subsequent paper by
Bak and Karch [3] found a more general solution of the matrix model describing an elliptical supertube, which includes a plane parallel D2/anti-D2 pair as a limiting case.   Mateos, Ng and Townsend had finally found supertubes with arbitrary cross-section [4] 

   The original supertube carries two independent conserved charges (D0 and F1).  Besides their intrinsic interest, supertubes are beginning to play an important role in black hole physics, based on the work of  Mathur and Lunin [5].  After a chain of dualities, the various configurations of the two charge supertubes are in one-to-one correspondence with the supersymmetric ground states of the D1-D5 system (with vanishing momentum).   The black hole constructed in there is different from those constructed by Strominger and Vafa [6].

  In an interesting paper  Lunin, Maldacena, and Maoz [7] conjectured that the entropy of a supertube configuration with $Q_{D0}$ units of D0 charge and $Q_{F1}$ units of F1 charge  is $S \sim \sqrt{Q_{D0}Q_{F1}}$ to leading order in large charges.  It is also of interest to count supertubes with fixed angular momentum $J$, in which case the corresponding conjecture of [8] would be
$$S \sim \sqrt{ (Q_{D0}Q_{F1}-J)}.  \eqno{(1.1)}$$ 
   In paper [9] Palmer and Marolf  counted the quantum states of the supertube by directly quantizing the linearized Born-Infeld action near the round tube.  In paper [10]  Bak et. al. counted the geometrically allowed microstates with fixed conserved charges and angular momenta in two different approaches using the DBI action and the supermembrane theory.   The results they found are consistent with the conjecture of [8], i.e. Eq.(1.1).

   Supertubes could also be described by the tachyon field theory.   In [11]  Kim et. al. investigated the supertube from the Dirac-Born-Infeld tachyonic action [12,13], in the spirit of the Sen's conjecture that the BPS branes can be viewed as tachyon kinks on non-BPS branes on higher dimension [13,14] - the remarkable `Decent Relation'.   The nontrivial coaxial array of tubular solution they found is the bound state of fundamental strings, D0-branes, and a cylindrical D2-brane and exhibits BPS-like property.  In paper [15] we used the Minahan-Zwiebach tachyon action [16] to find a single regular tube solution with circular and elliptic cross section. We had also calculate the fluctuation spectrum around the kink solution and see that there is no tachyonic mode therein.  These results are consistent with the identification of the tubular configuration or  as a BPS D2-brane. 

   In this paper we will present a possible way to count the quantum states of the supertube by directly quantizing the deformed modes of tachyonic Dirac Born-Infeld action near the round tube.    As a tube with slight deformation from a round tube possesses an arbitrary cross section, we shall first show that there are tachyon supertubes with arbitrary cross section and check up their supersymmetric property.  To do this we will in section II use the Minahan-Zwiebach tachyon action to find the single regular tube solution with arbitrary cross section.   We will see that the energy of the single tubular configuration comes entirely from the D0 and strings at critical Born-Infeld (BI) electric field.  This indicates that the solution is supersymmetric [1,4].  We will also calculate the fluctuation spectrum around the kink solution and show that there is no tachyonic mode therein.  These results are consistent with the identification of the tubular configuration as a stable BPS D2-brane.   

  In section III we will quantize the deformed modes of the  single circular supertube solution and use the result to count the allowed quantum states  in a supertube ensemble with fixed macroscopic quantities of charges $Q_{D0}$ ,$Q_{F1}$ and angular momentum $J$.   We see that the entropy of the system is proportional to $\sqrt{Q_{D0}Q_{F1} - J}$, which is consistent with that calculated by the DBI action [9,10].   Therefore, besides the well known properties that the kink solution (and its fluctuation) of tachyon DBI action corresponds with the brane solution (and its fluctuation) of DBI action, our result have established a new property that the entropy of the tachyon supertubes in tachyonic DBI action corresponds with that of the supertube in  DBI action.  We make a conclusion in the last section.

\section {Tube Solution in Minahan-Zwiebach Tachyon Action}
The Minahan-Zwiebach (MZ) tachyon action [16] is a derivative truncation of the BSFT action of the non-BPS branes [17], which embodies the tachyon dynamics for unstable D-branes in (super)string theories and was first proposed as a simplified action to capture the desirable properties of string theories.    The action was successfully used by Hashimoto and Nagaoka [18] to show the phenomena of kink condensation and vortex condensation in the unstable non BPS branes.   It supports the Sen's conjecture of the `Brane Descent Relations' of tachyon condensation.  We have also used the MZ tachyon action to investigate the problems of the interaction between the kink-anti-kink configurations and recombination of intersecting branes [19].    

\subsection{Tube with Arbitrary Cross Section}
The Minahan-Zwiebach tachyon action of  the non-BPS D3 brane is described by [16]
$$ S = -{\cal T}_3 \int dtdXdYdz\; V(T) \left(1+ (\partial_\mu T)^2 + {1\over 4} F_{\mu\nu}^2\right) . \eqno{(2.1)}$$
We will express the coordinates as 
$$X = r f(\theta)\, cos(\theta),  \eqno{(2.2a)}$$
$$Y = r f(\theta)\, sin(\theta),  \eqno{(2.2b)}$$
in which $0 \leq r < \infty$ and $0 \leq \theta < 2\pi$. The function $f(\theta)$ is an arbitrary function which determine the form of the closed curve on the x-y plane for a given value of $r$.   For example, in figure 1 we plot the curves for the cases of $f(\theta) = 1$ (which is a circle) and $f(\theta) = {1 + 0.2 \,sin(6\theta)}$ with $r=1$.
\\
\\
\scalebox{1}{\hspace{4cm}\includegraphics{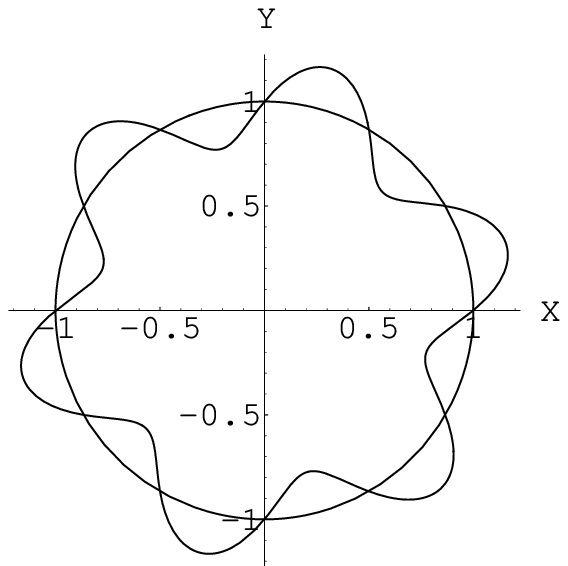}}\\ 
\hspace{2cm}{\it ~~~Figure.1. The curves for the cases of $f(\theta) = 1$ (which is a circle) and $f(\theta) = {1 + \epsilon \,sin(6\theta)}$ with $r=1$ and $\epsilon = 0.2$.  The later case may be regarded as a slightly deformed curve from a circle if $\epsilon \ll1$.}\\
\hspace{1cm}
The BI electric and magnetic fields are taken to be 
$$ F_{tz} = E,~~~F_{\theta z} = B,\eqno{(2.3)}$$
with constant $E$ and $B$.  Other components of EM field strength are vanishing.  In this case, the tachyonic effective action simplifies drastically
$$S =  -{\cal T}_3\int dt dz d\theta \int dr V(T) \, r\, f(\theta)^2 \left[\left(1-{1\over2}E^2\right) + {B^2\over 2r^2 f(\theta)^2} + {f(\theta)^2 + \dot {f}(\theta)^2\over f(\theta)^4} T'(r)^2 \right] ,\eqno{(2.4)}$$
where the prime ${}'$ (dot  $\dot {}$\,\,)denotes differentiation with respect to the radial coordinate $r$ (polar angle $\theta$).  Note that, although we let the tachyon field $T(r)$ merely depend on the radial coordinate $r$ the cross section of the tachyon tube solution $T(r)_c$ found in (2.8) could  have any curved form because $f(\theta)$ is an arbitrary function (figure 1 is just a simple example).
   
  After the integration over $\theta$ the  action (2.4) becomes
$$S =  -{\cal T}_3\int dt dz \int dr V(T) \, r\, \left[\alpha\left(1-{1\over2}E^2\right) + {B^2\over 2r^2 } + \gamma \,T'^2  \right] ,\eqno{(2.5)}$$
where
$$\alpha = \int _0^{2\pi} d\theta f(\theta)^2, ~~~~~~~~~~~~\eqno{(2.6a)}$$
$$\gamma = \int _0^{2\pi}  d\theta \,{f(\theta)^2 + \dot {f}(\theta)^2\over f(\theta)^2}. \eqno{(2.6b)}$$
\\
As the action (2.5) becomes that of round tube analyzed in our previous paper [15] if $\alpha = \gamma = 2 \pi$, we can therefore adopt the previous method to perform the following analyses. 

  The associated equation of motion form the action (2.5) is 
$$2V(T) \left(T''(x) + {T'\over r}\right) - V'(T)\left[\alpha\,\left(1-{E^2\over2}\right)+ {B^2\over 2r^2} - \gamma\,T'^2 \right] = 0, \eqno{(2.7)}$$
which can be solved by 
$$T(r)_c = {~B\over {\sqrt {2 \gamma}}}\, ln(r/r_0), \eqno{(2.8)}$$
if electric $E=E_c \equiv \sqrt 2$.  The value of  $r_0$ in above is an arbitrary integration constant.   As the value of $|T(r)_c|$ become zero at $r=r_0$ the radius of tube will depend on the value of  $r_0$.   The tube radius and $r_0$ is determined by the BI EM field, i.e., the charges of D0 and strings on the brane [15].   It is worthy to mention that the function form of the tachyon solution is irrelevant to  the function form of the tachyon potential $V(T)$, as those in the other tachyon kink solutions [15,18].

    To proceed, we define the electric displacement defined by $\Pi = \partial L/\partial E$ and thus from (2.5) 
$$\Pi =  {\cal T}_3 V(T) \, r\alpha \,E ,\eqno{(2.9)}$$
The associated energy density defined by $H = \Pi E - L$  becomes
$$H  =  {\cal T}_3 V(T) \,r \left[\alpha \left(1+{1\over2}E^2\right) + {B^2\over  r^2} \right] = {\cal T}_3 V(T) \,r \left(2\,\alpha + {B^2\over  r^2} \right) ,\eqno{(2.10)}$$
when $E=E_c$.  In figure 2 we plot the typical behaviors of function $H(r)$ which shows that there is a peak at finite radius.
\\
\\
\scalebox{1}{\hspace{4cm}\includegraphics{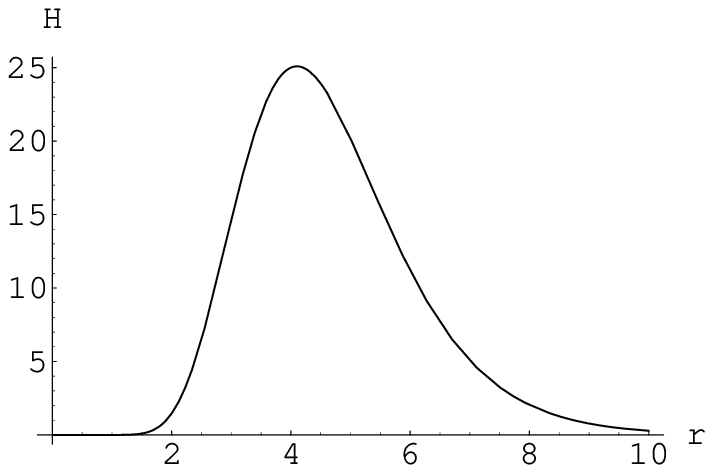}}\\ 
\\
\hspace{2cm}{\it ~~~Figure.2. The function behaviors of  $H(r)$ in (2.10) for the case of $B=6, \gamma = 3, \alpha =2 \,\, and \,\, r_0=4$.  There is a peak at finite radius r, which specifies the size of the tube cross section.}\\

   It is important to mention the physical meaning of the critical value of electric field $E_c$ [20].  The action (2.3) tell us that the electric field has the effect of reducing the brane tension, and increasing $E$ to its `critical' value $E_c=\sqrt 2$ would reduce the tension to zero if the magnetic field were zero; because that ${\cal L} \sim  B^2$ when $E=E_c$ (Note that from (2.8) we have a relation ${T'_c}^2 \sim B^2$).   This implies that the tachyon tube has no energy associated to the tubular D2-brane tension; its energy comes entirely from the electric and magnetic fields, which can be interpreted as `dissolved' strings and D0-branes, respectively. The energy from the D2-brane tension has been canceled by the binding energy released as the strings and D0-branes are dissolved by the D2-brane.  The phenomena that the tubular D2-brane tensor has been canceled was crucial to have a supersymmetric tube configuration [1,4,20]. 

\subsection{Fluctuation around Tubular Kink}
Let us consider the tachyon field $T(r)$ with fluctuation $\hat{T}(r)$ around the tubular solution $T(r)_c$, i.e.
$$T(r) = T(r)_c + \hat{T}(r).  \eqno{(2.11)}$$
Substituting the tubular kink solution (2.10) into the action (2.5) and considering only the quadratic terms of fluctuation field $\hat{T}(r)$ we obtain 
$$S = - {\cal T}_3 \,\gamma\,\int dr\,  \left( r\, \hat{t'}^2\, + \left( -{B^2\over 2r} + {B^4\over 4r} \,\,\left(ln\left({r/ r_0}\right) \right)^2\right) \hat{t}^2 \right). \hspace{2.2cm}\eqno{(2.12)}$$
in which we have used an partial integration and the field redefinition 
$$\hat{t} \equiv V(T_c \,\gamma)^{1\over2} \,\hat{T}(r),~~~~~~~~~with ~~~~V(T_c) = e^{-T_c^2}. \eqno{(2.13)}$$
The action (2.12) becomes that of round tube analyzed in our previous paper [15] if $ \gamma = 2 \pi$.   Defining a new variable 
$$w= ln\left({r/ r_0}\right),\eqno{(2.14)}$$
we can obtain
$$S =- 2\pi\,{\cal T}_3 \,\gamma\,\int dw\, \left[ \partial_w\hat{t}\, \partial_w\hat{t} + \hat{t}\left( -{B^2\over 2} + {B^4\over 4} \, w^2  \right)\hat{t}\right]. \eqno{(2.15)}$$
From the above expression we see that the fluctuation $t$ obeys a Schr\"odinger equation of a harmonic oscillator, thus the mass squared for the fluctuation is equally spaced and specified by an integer $n$, 
$$m^2_t=  2n\,B^2 , ~~~~~ \ n \geq 0 .\eqno{(2.16)}$$
Thus there is no tachyonic fluctuation, the mass tower starts from a massless state and has the equal spacing. This result is consistent with the identification of the tachyon tube of {\it arbitrary cross section}  as a tubular BPS D2-brane. 

    In conclusion, we have found the tachyon tube solution with arbitrary cross section and from its energy form and associated fluctuation we see that the solution is consistent with the identification of the tachyon tube as a tubular BPS D2-brane.  

   We will in the next section examine the ensemble of supertubes with deformed curved cross section from a circular supertube.  We will quantize the deformed modes and show that the entropy for the allowed quantum states  in a supertube ensemble with fixed macroscopic quantities of charges $Q_{D0}$ ,$Q_{F1}$ and angular momentum $J$ is proportional to $\sqrt{Q_{D0}Q_{F1} - J}$, which is consistent with that calculated by the DBI action [9,10].

\section{Quantum States of Supertube}
\subsection{Quantization of Deformed Modes}
To consider the quantum states of supertube deformed from that with circular cross section we will consider the following tachyon field 
$$T(t, r, \theta) = T(r)_c + \hat{T}(t, \theta).  \eqno{(3.1)}$$
in which $ T(r)_c$ is the circle tachyon supertube solution and $\hat{T}(t,\theta)$ is the fluctuation around it.   As the fluctuation field $\hat{T}(t,\theta)$ depends on the angle $\theta$ the tachyon field $T(t, r,\theta)$ may be used to described a tachyon supertube deformed from a one with circular cross section.  (Note that we have in section II found the tachyon tube solution with arbitrary cross section and shown that the solution is a stable tubular BPS D2-brane in the case of critical electric $E = E_c$.)

   Substituting the tachyon field (3.1) into the action (2.4) and considering only  the quadratic terms of fluctuation field we obtain 
$$S = - {\cal T}_3 \,\int  dt dr\,d\theta\, r \left[ V(T_c)\left(- \left(\partial_t\, \hat{T}(t,\theta)\right)^2 + {1\over r^2}\left(\partial_\theta\, \hat{T}(t,\theta)\right)^2 \right) + {B^2\over 2r^2} {V''(T_c) \over 2}\, \hat{T}(t,\theta)^2 \right]$$
$$ =  {\cal T}_3 \,\int  dt d\theta \left[{\sqrt {2\pi}\over B} \, r_0^2\,\left(\partial_t\, \hat{T}(t,\theta)\right)^2 -  {\sqrt {2\pi}\over B} \left(\partial_\theta\, \hat{T}(t,\theta)\right)^2 \right], ~~~~~~~~~~~~~~~~~~~~~~~~\eqno{(3.2)}$$
in which as the background is the circle tachyon supertube we shall let $f(\theta) =1$ in (2.4).  We have also used the relation $V(T) = e^{-T^2}$, and to simplify the relation we assume that $B \gg 1$ (More precisely, we approximate $e^{2/B^2} \approx 1$). 

   The field equation of  $\hat{T}(t,\theta)$ read from above equation is 
$$ r_0^2\,\partial_t^2\, \hat{T}(t,\theta) - \partial_\theta^2\, \hat{T}(t,\theta) = 0,\eqno{(3.3)}$$ 
and the mode expansion of  $\hat{T}(t,\theta)$ becomes
$$\hat{T}(t,\theta) = \sum_n a_n e^{i\omega_n \, t- i n \theta},\eqno{(3.4)}$$ 
in which the relation 
$$\omega_n = n/r_0, \eqno{(3.5)}$$
could be read from field equation (3.3).

   The conjugate momentum of fluctuation field $\hat{T}(t,,\theta)$ is 
$$\Pi_{\hat{T}} = 2 {\cal T}_3 \, r_0^2\,{\sqrt {2\pi}\over B} \,\partial_t\, \hat{T}(t,\theta). \eqno{(3.6)}$$
Then, after imposing the quantization relation
$$[\,\Pi_{\hat{T}}(t,\theta), \hat{T}(t,\tilde \theta)\,] = - i \delta\left(\theta-\tilde \theta\right), \eqno{(3.7)}$$
we have the following commutative relation 
$$ [ a_n, a_m] =  \delta_{n,-m}{B\over \sqrt{2\pi}\, 2n \,{\cal T}_3 \, r_0}.\eqno{(3.8)}$$
In terms of 
$$c_n \equiv  \left(\sqrt {B\over \sqrt{2\pi}\, 2n {\cal T}_3 \, r_0 }\,\right)^{-1}\,a_n  , ~~~~~c_n^\dag \equiv \, c_{-n}, \eqno{(3.9)}$$
\\
we have a simple commutation relation 
$$[\,c_n , c_n^\dag\,]= \delta _{m, n} . \eqno{(3.10)}$$
In next subsection we will express the quantity  $Q_{D0}Q_{F1} -J$  in terms of the mode operators $c_n$   and see that the entropy of the tube ensemble we consider  is proportional to $\sqrt{Q_{D0}Q_{F1} - J}$.

\subsection{Entropy of Tachyon Supertube}
To proceed  we can from the conventional definitions to calculate the following relations
$$J = {{\cal T}_3 }\int dr\,d\theta \, \, r\left[V(T_c) + V'(T_c) \,\hat{T}(t, \theta)+ {1\over 2}V''(T_c) \, \hat{T}(t, \theta)^2 \right] \, E_c \, B ~~ $$
$$ = {\cal T}_3 \, 4 \pi \sqrt \pi \left[1 + {8\over B^2} \sum_n a_n^\dag a_n \right] \, r_0^2\equiv J^{(0)}+ \Delta J. \hspace{3cm}\eqno{(3.11a)}$$

 $$\tilde{Q}_{F1} = {{\cal T}_3 }\int dr\,d\theta \, \, r\left[V(T_c) + V'(T_c)\, \hat{T}(t, \theta)+ {1\over 2}V''(T_c)  \,\hat{T}(t, \theta)^2 \right] \, E_c \,  $$
$$ = {\cal T}_3 \, {4 \pi \sqrt \pi \over B}\left[1 + {8\over B^2} \sum_n a_n^\dag a_n \right] r_0^2 \equiv \tilde{Q}_{F1}^{(0)}+ \Delta \tilde{Q}_{F1}. \hspace{1.5cm}\eqno{(3.11b)}$$

$$\tilde{Q}_{D0} = {{\cal T}_3 }\int dr\,d\theta \,\, r\left[\left(V(T_c) + V'(T_c) \,\hat{T}(t, \theta)+ {1\over 2}V''(T_c) \, \hat{T}(t, \theta)^2 \right) {B^2\over  r^2} + V(T_c) \right.$$
$$\left. ~~~~~~~~~ \times \left({1\over r^2}\left(\partial_\theta^2\, \hat{T}(t,\theta)\right)^2  + \left(\partial_ t^2\, \hat{T}(t,\theta)\right)^2\right)\right] = {\cal T}_3 \, {2 \pi \sqrt 2 \pi \over B}\left[B^2  +  \sum_n n ^2 a_n^\dag a_n \right]~~~~~$$

$$  \equiv \tilde{Q}_{D0}^{(0)}+ \Delta \tilde{Q}_{D0},  \hspace{8.5cm}\eqno{(3.11c)}$$
in which the $\Delta$-terms are those contain operator $ a^\dag_n a_n$.  The definition of $\tilde{Q}_{D0} $ in above will satisfy the energy relation  $U = \tilde{Q}_{D0}  + \tilde{Q}_{F1} $, which is a property indicating that the deformed tube is a supersymmetric configuration.  

   We can now follow [4] and [9] to define the normalized charges:
$$ {Q}_{F1} = {\tilde{Q}_{F1} \over {\cal T}_1 }  \equiv {Q}_{F1}^{(0)}+ \Delta {Q}_{F1},\eqno{(3.12a)}$$
$$ {Q}_{D0} = \sqrt {2\pi}\, \,{\tilde{Q}_{D0} \over {\cal T}_0 }  \equiv {Q}_{D0}^{(0)}+ \Delta {Q}_{D0} .\eqno{(3.12b)}$$
Therefore, without the deformation the circular supertube will have the charges $Q^{(0)}_{D0}$ , $Q^{(0)}_{F1}$ and angular momentum $J^{(0)}$, which satisfy the relation [1]
$$Q^{(0)}_{D0}~~Q^{(0)}_{F1} = J^{(0)}, \eqno{(3.13)}$$
as can be seen from (3.11) and the definition of ${\cal T}_p = {2\pi\over (2\pi \l_s)^{p+1}} $, with the unit $\l_s=1$.   Finally, from (3.11) and (3.12) we have a simple expression 
$$  \left(Q_{D0}Q_{F1} - J\right) \sim \sum_n \, n c_n^\dag \, c_n = \sum_n \, n \, N_n,\eqno{(3.14)}$$
in which $N_n$ is the number operator of $c_n$.

   The remained work is to evaluate the number of quantum states $\Omega $ restricted by the relation (3.14).  As the problem is the same as the well-known case of counting string states [21],  the entropy $S$ of the ensemble of supertube with fixed macroscopic quantities of charges $Q_{D0}$ ,$Q_{F1}$ and angular momentum $J$  is then given by 
$$ S = Log(\Omega) \sim  \sqrt {Q_{D0}Q_{F1} - J},\eqno{(3.15)}$$
as claimed in the introduction.

  Let us make following comments to conclude this section.

1.  The relation (3.12b) has a factor $\sqrt {2\pi}$ which does not show in [9].  It is a consequence of the particular tachyon potential we adopted. (Note that the tube solution in (2.8) is irrelevant to the tachyon potential.)

2.  Our investigations are based on the single regular tube solution of MZ tachyon action.  The authors in [22] had found a single thin tube solution.  It is interesting to calculate the tube entropy of their solution.

3.  The tachyon field considered in (3.1) is $T(t,r,\theta) = T(r)_c + \hat{T}(t, \theta)$ and the associated entropy of Eq.(3.15) is only that from a single quantized fluctuating boson operator  $\hat{T}(t, \theta)$.  On the general grounds one shall use the tachyon field $T(t, x_1,.., x_8)$, in which $x_i \not= z$ as the tube  is along the $z$-axial and it has a fixed angular momentum $J=J_z$.
After expanding the tachyon field $T(t, x_1,.., x_8)$ around the tube solution $T(r)_c$  there will have eight fluctuation fields  $\hat{T}(t, x_i)$ and each of them will contribute to the tube entropy.  Also, as the tachyon tube is an 1/4 supersymmetric configuration [1], besides the $8$ classes of bosonic number operators there will have  $4$ classes of fermionic number operators.   Thus the total entropy is $\sqrt{c_B+c_F}=\sqrt{12}$ times the entropy S calculated merely from the bosonic fluctuation field $\hat{T}(t, \theta)$.   This property had been checked by Bak. et. al. in the paper [10] by using the DBI action.   The checking work in the tachyonic DBI action will be more involved and remains to be calculated.

\section{Conclusion}
In this paper we have extended previous paper [15] to find tachyon supertubes with arbitrary cross section and see that the energy of the single tubular configuration comes entirely from the D0 and strings at critical Born-Infeld (BI) electric and magnetic fields.  This indicate that the solution is supersymmetric.    We also have calculated the fluctuation spectrum around the kink solution and shown that there is no tachyonic mode therein.  These results are consistent with the identification of the tubular configuration as a stable BPS D2-brane.   We have quantized the deformed modes of a supertube solution with regular profile of circular cross section on the unstable non-BPS D3-branes by using the Minahan-Zwiebach tachyon action.  We have counted the allowed quantum states in an ensemble of supertube with fixed macroscopic quantities of charges $Q_{D0}$ ,$Q_{F1}$ and angular momentum $J$.   We finally show that the entropy of the system is proportional to $\sqrt{Q_{D0}Q_{F1} - J}$, which is consistent with that calculated by the DBI action.  Our result establishes a new property that the entropy of the tachyon supertubes in tachyonic DBI action corresponds with that of the supertube in  DBI action.

  Finally we want to mention two interesting works.  
\\
1. We have shown that the tachyon tube (which is described by a tachyon field) has the same entropy as that in brane tube (which is described by the coordinates in DBI action).  As the radiation and decay of black hole have been investigated in brane picture [6] it is therefore worthy to study the radiation or decay of the tube (or black hole) in the tachyon field theory.
\\
2. The recent investigations [23] have shown a possible way to investigate the Schwarzschild black hole in brane-antibrane systems.  As the brane-antibrane system is a non-BPS state it can be described by the tachyon field. The neutral black hole may therefore be studied by tachyon field theory.

The interesting problems remain to be studied.
   
\newpage
 
\begin{center} {\large \bf  References} \end{center}
\begin{enumerate}
\item D. Mateos and P. K. Townsend, ``Supertubes'', Phys. Rev. Lett. 87 (2001) 011602 [hep-th/0103030]; R. Emparan, D. Mateos and P. K. Townsend, ``Supergravity Supertubes'', JHEP 0107 (2001) 011 [hep-th/0106012].
\item D. Bak, K. M. Lee, ``Noncommutative Supersymmetric Tubes'',  Phys. Lett. B509 (2001) 168 [hep-th/0103148];D. Bak and S. W. Kim, ``Junction of Supersymmetric Tubes,'' Nucl. Phys.  B622 (2002) 95 [hep-th/0108207].
\item  D. Bak and A. Karch, ``Supersymmetric Brane-Antibrane Configurations,'' Nucl. Phys.  B626 (2002) 165 [hep-th/011039]; D. Bak and N. Ohta, ``Supersymmetric D2-anti-D2 String,'' Phys. Lett.  B527 (2002) 131 [hep-th/0112034]; D. Bak, N. Ohta and M. M. Sheikh-Jabbari, ``Supersymmetric Brane-Antibrane Systems: Matrix Model Description, Stability and Decoupling Limits,'' JHEP  0209 (2002) 048 [hep-th/0205265].
\item  D.~Mateos, S.~Ng and P.~K.~Townsend, ``Tachyons, supertubes and brane/anti-brane systems'', JHEP  0203 (2002) 016 [hep-th/0112054]; Y. Hyakutake and N. Ohta, ``Supertubes and Supercurves from M-Ribbons,'' 
Phys. Lett. B539 (2002) 153 [hep-th/0204161]. 
\item  O.~Lunin and S.~D.~Mathur, ``Metric of the multiply wound rotating string,'' Nucl.\ Phys.\ B \ 610 (2001) 49 [hep-th/0105136]; O.~Lunin and S.~D.~Mathur, ``AdS/CFT duality and the black hole information paradox,''
Nucl.\ Phys.\ B  623(2002) 342 [hep-th/0109154]; S.~D.~Mathur, A.~Saxena and Y.~K.~Srivastava, ``Constructing 'hair' for the three charge hole,'' Nucl.\ Phys.\ B \ 680 (2004) 415 [hep-th/0311092]; I. Bena and P. Kraus, ``Three Charge Supertubes and Black Hole Hair'' , Phys.Rev. D70 (2004) 046003 [hep-th/0402144].
\item A.~Strominger and C.~Vafa, ``Microscopic origin of the Bekenstein-Hawking entropy,'' Phys.\ Lett.\ B  379 (1996) 99 [hep-th/9601029]; C. G. Callan, J. M. Maldacena , ``D-brane Approach to Black Hole Quantum Mechanics,'' Nucl.Phys. B472 (1996) 591[hep-th/9602043]; J. R. David, G. Mandal, S. R. Wadia, ``Microscopic Formulation of Black Holes in String Theory,'' Phys.Rept. 369 (2002) 549 [hep-th/0203048].
\item  O.~Lunin, J.~Maldacena and L.~Maoz, ``Gravity solutions for the D1-D5 system with angular momentum,'' [hep-th/0212210].
\item  O.~Lunin and S.~D.~Mathur,  ``Statistical interpretation of Bekenstein entropy for systems with a stretched horizon,'' Phys.\ Rev.\ Lett.\ 88 (2002) 211303 [hep-th/0202072].
\item B. C. Palmer and D. Marolf , `` Counting Supertubes'',  JHEP 0406 (2004) 028 [hep-th/0403025].
\item   D. Bak, Y. Hyakutake, S. Kim and N. Ohta, ``A Geometric Look on the Microstaties of Supertubes,'' [hep-th/0407253]; D. Bak, Y. Hyakutake, and N. Ohta, `` Phase Moduli Space of Supertubes,'' [hep-th/0404104]
\item C. Kim, Y. Kim, O-K. Kwon, and P. Yi, ``Tachyon Tube and Supertube,''  JHEP 0309 (2003) 042 [hep-th/0307184].
\item M. R. Garousi, `` Tachyon couplings on non-BPS D-branes and Dirac-Born-Infeld action,''  Nucl.Phys. B584 (2000) 284 [hep-th/0003122].
\item A. Sen, ``Tachyon Condensation on the Brane Antibrane System'', JHEP 9808 (1998) 012, [hep-th/9805170]; ``Descent Relations Among Bosonic D-branes'',  Int.\ J.\ Mod.\ Phys. A14 (1999)  4061, [hep-th/9902105];  ``Non-BPS States and Branes in String Theory'',  [hep-th/9904207];  ``Universality of the Tachyon Potential'',  JHEP 9912 (1999) 027, [hep-th/9911116]; ``Supersymmetric world-volume action for non-BPS D-branes,''  JHEP  9910 (1999) 008 [hep-th/9909062].
\item A. Sen, ``Dirac-Born-Infeld Action on the Tachyon Kink and Vortex,''  Phys. Rev. D68 (2003) 066008 [hep-th/0303057].
\item Wung-Hong Huang ``Tachyon Tube on non-BPS D-branes,''  JHEP 0408 (2004) 060 [hep-th/0407081].
\item  J.  A. Minahan and B.  Zwiebach,``Gauge Fields and Fermions in Tachyon Effective Field Theories'',  JHEP 0102 (2001) 034 [hep-th/0011226]; ``Effective Tachyon Dynamics in Superstring Theory'',  JHEP  0103 (2001) 038 [hep-th/0009246].
\item  B.  Zwiebach, ``A Solvable Toy Model for Tachyon Condensation in String Field Thoery' , JHEP 0009 (2000) 028 [hep-th/0008227];  J.  A.  Minahan and B.  Zwiebach, ``Field Theory Models for Tachyon and Gauge Field String
Dynamics'',   JHEP 0009 (2000)  029 [hep-th/0008231]; O.   Andreev, ``Some Computations of Partition Functions and Tachyon Potentials in Background Independent Off-Shell String Theory'' ,  Nucl.Phys. B598 (2001) 151 [hep-th/0010218]; D. Kutasov and V. Niarchos, ``Tachyon effective actions in open string theory,''  [hep-th/0304045].
\item K. Hashimoto and S. Nagaoka,``Realization of Brane Descent Relations 
in Effective Theories'', Phys. Rev. D66 (2002) 0206001  [hep-th/0202079].
\item Wung-Hong Huang, ``Brane-Antibrane Systems Interaction under Tachyon Condensation'' , Phys.Lett. B561 (2003) 153 [hep-th/0211127]; Wung-Hong Huang, ``Recombination of Intersecting D-branes in Tachyon Field Theory'' , Phys.Lett. B564 (2003) 155 [hep-th/0304171].
\item P. K. Townsend, ``Surprises with Angular Momentum'', Annales Henri Poincare 4 (2003) S183 [hep-th/0211008].
\item M. B. Green, J. H. Schwarz and E. Witten, Superstring theory,
Cambridge University Press, 1987.
\item  L. Martucci and P. J. Silva, "Kinky D-branes and straight strings of open string tachyon effective theory",   JHEP 0308 (2000)  026 [hep-th/0306295].
\item U. H. Danielsson, A. Guijosa, and M. Kruczenski, ``Brane-Antibrane Systems at Finite Temperature and the Entropy of Black Branes'',  JHEP 0109 (2001) 011 [hep-th/0106201]; O. Saremi, A. W. Peet, ``Brane-antibrane systems and the thermal life of neutral black holes'',  Phys.Rev. D70 (2004) 26008 [hep-th/0403170]; O. Bergman, G. Lifschytz, ``Schwarzschild black branes from unstable D-branes'',   JHEP 0404 (2004) 060[hep-th/0403189].
\end{enumerate}
\end{document}